\documentclass[conference]{IEEEtran}
\IEEEoverridecommandlockouts

\usepackage{cite}
\usepackage{amsmath,amssymb,amsfonts}
\usepackage{algorithmic}
\usepackage{textcomp}
\usepackage{xcolor}
\usepackage{lipsum}
\usepackage{placeins}
\usepackage{multirow}
\usepackage{makecell}
\usepackage{graphicx}
\usepackage{subcaption}
\usepackage{booktabs}
\usepackage{hyperref}
\def\BibTeX{{\rm B\kern-.05em{\sc i\kern-.025em b}\kern-.08em
    T\kern-.1667em\lower.7ex\hbox{E}\kern-.125emX}}
\begin{document}

\title{Reading Radio from Camera: Visually-Grounded, Lightweight, and Interpretable RSSI Prediction
}

\author{
\IEEEauthorblockN{
Sen Yan,
Tianyu Hu\IEEEauthorrefmark{1},
Brahim Mefgouda,
Samson Lasaulce\IEEEauthorrefmark{2}
and Merouane Debbah
}

\IEEEauthorblockA{
\IEEEauthorrefmark{0}\textit{Computer and Information Engineering, Khalifa University}, Abu Dhabi, UAE
}

\IEEEauthorblockA{
\IEEEauthorrefmark{1}\textit{University of Electronic Science and Technology of China (UESTC)}, China
}

\IEEEauthorblockA{
\IEEEauthorrefmark{2}\textit{CNRS and Université de Lorraine, CRAN}, France
}

\IEEEauthorblockA{
Emails: yansen0508@gmail.com, huty@std.uestc.edu.cn,
brahim.mefgouda@ku.ac.ae, \\
samson.lasaulce@univ-lorraine.fr,
merouane.debbah@ku.ac.ae
}
}

\maketitle

\begin{abstract}
Accurate, real-time wireless signal prediction is essential for next-generation networks. However, existing vision-based frameworks often rely on computationally intensive models and are also sensitive to environmental interference. 
To overcome these limitations, we propose a novel, physics-guided and light-weighted framework that predicts the received signal strength indicator (RSSI) from camera images.  
By decomposing RSSI into its physically interpretable components, path loss and shadow fading, we significantly reduce the model's learning difficulty and exhibit interpretability.
Our approach establishes a new state-of-the-art by demonstrating exceptional robustness to environmental interference, a critical flaw in prior work. Quantitatively, our model reduces the prediction root mean squared error (RMSE) by 50.3\% under conventional conditions and still achieves an 11.5\% lower RMSE than the previous benchmark's interference-eliminated results.
This superior performance is achieved with a remarkably lightweight framework, utilizing a MobileNet-based model up to 19 times smaller than competing solutions. 
The combination of high accuracy, robustness to interference, and computational efficiency makes our framework highly suitable for real-time, on-device deployment in edge devices, paving the way for more intelligent and reliable wireless communication systems.

\end{abstract}

\begin{IEEEkeywords}
Received signal strength indicator prediction, path loss, shadow fading, computer vision, convolutional neural network
\end{IEEEkeywords}

\section{Introduction}
Accurate prediction of the received signal strength indicator (RSSI) is crucial for the design and optimization of next-generation wireless networks \cite{zhang2025vision,arxiv_2025_rgb}. Traditionally, this task relies on analytical signal propagation models, which offer clear physical interpretability but require precise environmental parameters and complex ray tracing that are often unavailable in real-world deployments. Consequently, their application is frequently confined to simulation or idealized scenarios. 
With the advent of deep learning, modern data-driven approaches can now predict RSSI from multimodal inputs like RGB images, visual semantics and GPS coordinates~\cite{jsac_depth_2020,wcnc_2024_smartfactory,arxiv_2025_rgb,isape_2024_cv,zhang2025vision}. 
However, these methods typically function as "black boxes," attempting a brute-force, end-to-end prediction of the final RSSI value. 
This approach not only obscures the underlying physical factors but also frames the problem as a complex, monolithic learning task. 
Therefore, such models lack physical interpretability and often exhibit poor robustness to the environmental interference.
Drawing inspiration from computer vision, where simplifying complex optimization tasks has led to performance breakthroughs (e.g., the shift from GANs \cite{gan,IMGA} to diffusion models \cite{ddpm,diffnmr1,diffnmr3}), \textbf{we pose a foundational question:} \textit{can the brute-force task of RSSI prediction be decomposed into simpler, more physically meaningful sub-problems?}

This paper answers this question in the affirmative by introducing a novel, physically interpretable framework for RSSI prediction. 
Instead of a brute-force regression, our model leverages a fundamental principle of wireless propagation: the decomposition of RSSI into its primary components. We explicitly predict path loss (PL), which captures distance-based attenuation, and shadow fading (SH), which models obstacle-induced fluctuations, directly from visual semantics. 
These physically meaningful components are then combined according to propagation theory to construct the final RSSI prediction. This structured, physics-guided approach yields our model that is not only interpretable but also establishes a new state-of-the-art in performance, efficiency, and robustness. 

The main contributions of this work are summarized as follows:
\begin{itemize}
\item \textbf{Physics-Guided Interpretable Learning.} We propose a novel framework that, guided by fundamental physical principles, decomposes the complex task of direct RSSI prediction into two simpler, physically interpretable sub-tasks: predicting PL and SH. By aligning the learning objective with the underlying physics of signal propagation, we significantly reduce the model's learning difficulty.

\item \textbf{State-Of-The-Art Performance with Superior Efficiency.} Our proposed model establishes a new SOTA in RSSI prediction. Critically, this is achieved with a remarkably lightweight framework, utilizing up to 19 times fewer parameters than competing models \cite{arxiv_2025_rgb}, highlighting its superior efficiency.

\item \textbf{Exceptional Robustness to Interference.} The model demonstrates exceptional robustness to environmental distractors. Its prediction error remains stable with almost no degradation in the presence of interference. This stands in stark contrast to the previous benchmark \cite{arxiv_2025_rgb}, which require manual removal of interference to function optimally and suffer significant performance drops otherwise.

\item \textbf{Potential for Edge Deployment:} By leveraging a small and efficient model, our work is suited for deployment on resource-constrained edge devices, such as in-vehicle computing systems and unmanned aerial vehicles. This combination of high accuracy, robustness, and low computational cost paves the way for future mobile networks.
\end{itemize}

\section{Related work}
In the context of predicting physically interpretable components of wireless signal propagation, prior works fall broadly into two major categories: (i) \textit{radio map-based RSSI prediction} using bird-view or abstract environment maps \cite{sato2017kriging,levie2021radiounet}, and (ii) \textit{camera-based RSSI prediction} using RGB or depth images. 
In the scope of this paper, our work is more relevant to the second category. The camera-based signal strength prediction can be further grouped into indoor and outdoor settings.

For indoor applications, the researchers utilize recurrent neural networks to predict RSSI from temporal sequences of depth images in static lab environments \cite{rnn_depth_2018}. The proactive approach in~\cite{jsac_depth_2020} uses depth sequences to anticipate power changes due to dynamic blockage in mmWave links. Privacy-preserving multimodal fusion is explored in~\cite{split_learning_2020}, where depth images are combined with RF features. Similarly,~\cite{wcnc_2024_smartfactory} applies vision-aided RSRP prediction to industrial factory settings using RGB images.

For outdoor scenes, the two-stage method proposed in~\cite{arxiv_2025_rgb} applies instance segmentation and object detection on RGB camera input to predict received power in vehicular communication settings. However, this approach functions as a "black box" that lacks physical interpretability. Furthermore, its two-stage architecture relies on massive models, making it unsuitable for resource-constrained edge devices.
The same team \cite{isape_2024_cv,zhang2025vision} proposes a vision-based framework to predict multiple physically meaningful channel characteristics, including large-scale fading, delay spread, and Rician $K$-factor, from RGB images of urban vehicular scenarios. The method leverages semantic segmentation to extract key scene elements (e.g., cars, pedestrians), followed by a supervised learning model to regress each parameter. 
While these works expand the prediction scope beyond RSSI, it does not explicitly decompose large-scale fading into interpretable components such as SH and PL and the prediction errors remain relatively large, i.e., 2 dB - 3 dB.

Although all the aforementioned works are pioneering, they highlight two persistent challenges in the field: the lack of a framework that is simultaneously interpretable, robust, and lightweight. A brute-force, non-interpretable prediction strategy not only hinders model reliability but also leads to massive parameter counts, which impedes deployment on edge devices. The motivation for our work stems precisely from our desire to address these critical gaps.

\section{Physically Interpretable Prediction of RSSI}
\subsection{Preliminaries: System Model}\label{proxy}
In this paper, we consider a communication scenario between a transmitter (Tx) and a receiver (Rx) in an urban environment. The focus is on predicting RSSI using camera images that capture both the Tx and Rx, along with their relative position. Notable, in wireless communication systems, the RSSI is fundamentally influenced by two major attenuation factors: PL and SH~\cite{hu20233d}. The relationship can be described in logarithmic scale as:

\begin{equation}
\text{RSSI} = P_t - \alpha - \beta,  
\label{eq:rssi}
\end{equation}where \( P_t \) is the transmit power, $\alpha$ denotes PL, the large-scale signal attenuation over distance.  $\beta$ represents SH, which accounts for additional \textcolor{black}{large-scale fading} fluctuations caused by obstacles obstructing the signal path. 
PL is typically modeled as:
\begin{equation}
    \alpha(n,d) = 10 \cdot n \cdot \log_{10}(d),
    \label{eq:pl}
\end{equation}where \( d \) is the distance between Tx and Rx, and \( n \) is the PL exponent that reflects the propagation environment (e.g., free space, urban, or suburban)~\cite{hu20233d}. 


Since the transmit power $P_t$ is unknown but usually stable, we cannot directly measure SH from the predicted PL based on \eqref{eq:pl}. Instead, we define a proxy term for SH by rearranging the equation: $\text{SH}:=P_t-\beta=\text{RSSI}+\alpha$.
Our model is trained to predict this combined term, which we will refer to as SH for simplicity throughout the rest of this paper. Consequently, the final RSSI is computed as:

\begin{equation}
    \text{RSSI} = -\text{PL} + \text{SH}.
    \label{eq:rssi_v2}
\end{equation}

\subsection{Model Architecture}
\begin{figure*}[t]
    \centering
    \includegraphics[width=0.95\textwidth]{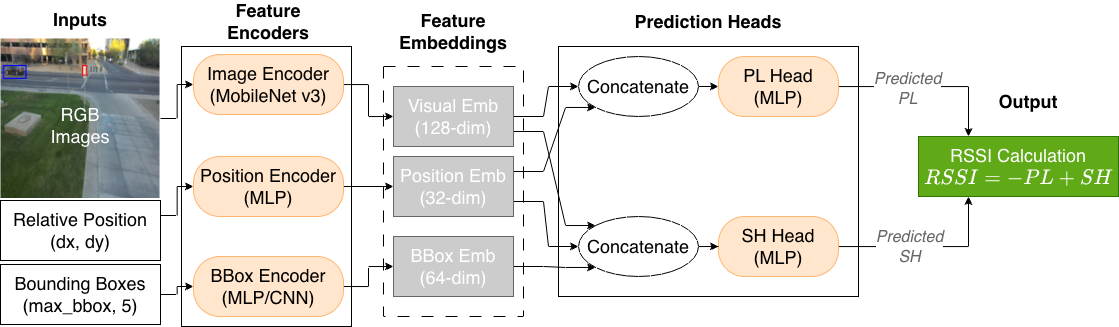}
    \caption{The framework of RSSI prediction based on real world images.}
    \label{fig:framework}
\end{figure*}
\begin{table*}[t]
\centering
\caption{Configuration of each module. For this work, $\text{max\_bbox}=10$. All layers are applied per sample (batch dimension omitted).}
\begin{tabular}{|l|c|c|c|c|}
\hline
\textbf{Module} & \textbf{Input Dim} & \textbf{Layer} & \textbf{Output Dim} & \textbf{Activation} \\
\hline
\multirow{2}{*}{BBox Encoder (MLP)} 
    & $\text{max\_bbox} \times 5$ & Linear & 128 & ReLU \\
    & 128 & Linear & 64 & None \\
\hline
\multirow{4}{*}{BBox Encoder (CNN)} 
    & $5 \times \text{max\_bbox}$ & Conv1d (k=1) & $32 \times \text{max\_bbox}$ & ReLU \\
    & $32 \times \text{max\_bbox}$ & Conv1d (k=1) & $128 \times \text{max\_bbox}$ & ReLU \\
    & $128 \times \text{max\_bbox}$ & AdaptiveAvgPool1d & 128 & None \\
    & 128 & Linear & 64 & None \\
\hline
Position Encoder (MLP)
    & 2 & Linear & 32 & ReLU \\
\hline
\multirow{2}{*}{PL Head (MLP)} 
    & 160 (=128+32) & Linear & 64 & ReLU \\
    & 64 & Linear & 1 & None \\
\hline
\multirow{2}{*}{SH Head (MLP)} 
    & 224 (=64+128+32) & Linear & 64 & ReLU \\
    & 64 & Linear & 1 & None \\
\hline
\end{tabular}
\label{tab:module-architectures}
\end{table*}
As illustrated in \autoref{fig:framework}, the framework is composed of several core components designed to process multimodal inputs and predict the physical constituents of RSSI in a structured manner. The framework consists of three parallel feature encoding branches, two specialized prediction heads for PL and SH, thus making final prediction based on \eqref{eq:rssi_v2}.

The model takes three distinct types of data as input:
\begin{itemize}
    \item An RGB image of the scene captured from the receiver's perspective.
    \item The 2-dimensional (2D) relative position vector $(dx, dy)$ between Tx and Rx, derived from GPS coordinates.
    \item Bounding box information for the Tx and up to 10 potential distractors (e.g., other vehicles, pedestrians), which encode object location, size, and class.
\end{itemize}

These inputs are processed by the following parallel encoders:

\begin{itemize}
    \item \textbf{Image Encoder:} A vision backbone is employed to extract high-level semantic features from the input RGB image. We select MobileNet \cite{howard2017mobilenets,mobilenet} for its proven efficiency and effectiveness, making it well-suited for potential deployment on edge devices.

    \item \textbf{Position Encoder:} Implemented as a multi-layer perceptron (MLP), this processes the 2D relative position vector to generate a compact positional embedding. The embedding helps the model learn the fundamental relationship between distance and path loss.

    \item \textbf{Bounding Box Encoder:} This module encodes the set of bounding box tensors. It is designed to capture the influence of occlusions from the Tx and surrounding objects, which is critical for accurately modeling shadow fading effects. For this encoder, both MLP and Convolutional Neural Network (CNN) architectures \cite{cnn} are explored in our experiments to determine the most effective representation.
\end{itemize}

The features from these encoders are then fused via concatenation to create combined representations for the two prediction heads:

\begin{itemize}
    \item \textbf{PL Head:} This head, implemented as an MLP, takes the fused features from the \textbf{Image Encoder} and \textbf{Position Encoder} as input. Its sole task is to regress the large-scale PL, leveraging both visual context and precise distance information.

    \item \textbf{SH Head:} This head, also an MLP, receives a richer set of features by fusing inputs from all three encoders: the \textbf{Image Encoder}, \textbf{Position Encoder}, and \textbf{Bounding Box Encoder}. This allows it to model the complex shadow fading effects caused by specific object occlusions in the scene. 
\end{itemize}

As a result, the RSSI is not directly regressed but is calculated using the physically grounded formula described in \eqref{eq:rssi_v2}, combining the outputs from the PL and SH heads. This structured approach ensures that the final prediction is both accurate and physically interpretable. The configuration of each module is detailed in \autoref{tab:module-architectures}.

\section{Experimental Setup}
\subsection{Dataset and Preprocessing}
For our experiments, we use the DeepSense 6G dataset \cite{deepsense}, a large-scale, multi-modal benchmark featuring real-world data for sensing and communication research. We utilize data from Scenarios 1 through 9, which encompass a variety of outdoor urban scenes under different conditions, such as day and night. The following data modalities are used as inputs and ground truth for our model:
\begin{itemize}
    \item \textbf{RGB Images:} Captured from a fixed camera perspective at 30 fps with a resolution of $960 \times 540$ pixels.
    \item \textbf{GPS Coordinates:} Provided for both Tx and Rx.
    \item \textbf{Bounding Boxes:} Annotations in YOLO format \cite{yolo} for the Tx and other dynamic objects (i.e. distractors).
    \item \textbf{RSSI Measurements:} Derived from a 60 GHz phased array receiver, consisting of 64 beams.
\end{itemize}

These raw data streams are preprocessed as follows:
\begin{itemize}
    \item \textbf{Images} are resized to $224 \times 224$ pixels and normalized to be compatible with the input dimensions of the pre-trained image encoder.
    \item \textbf{GPS Coordinates} are converted into a Min-Max normalized relative position vector $(dx, dy)$ and the real Euclidean distance $d$ between the Rx and Tx using the pyproj library~\cite{pyproj}. \enlargethispage{\baselineskip}
    \item \textbf{Bounding Box} tensors are formatted to represent up to 10 objects, capturing information about the Tx and potential obstructions.
    \item \textbf{Ground Truth RSSI} is calculated by taking the mean power across the 64 received beams and converting it to dB.
    \item \textbf{Ground Truth PL} is calculated using \eqref{eq:pl}. 
    For the open urban vehicle-to-infrastructure scenario in the dataset, we approximate the path loss exponent as $n \approx 2$. The distance $d$ is given via GPS coordinates.
    \item \textbf{Ground Truth SH} is obtained by computing the proxy term $P_t - SH = \text{RSSI} + \text{PL}$, as described in \autoref{proxy}.
\end{itemize}

\begin{table}[t]
\centering
\caption{The best of the two-stage training configurations. Note that BS = batch size; LR = learning rate; Sched. = Scheduler; W.U = warm-up; C.A = Cosine Annealing.}
\label{tab:training-config}
\begin{tabular}{lcccccc}
\toprule
\textbf{Stage} & \textbf{Epochs} & \textbf{BS} & \textbf{Optim.} & \textbf{LR} & \textbf{\makecell{LR \\ Sched.}} & \textbf{\makecell{$\lambda_{\text{PL}}$, \\ $\lambda_{\text{SH}}$}} \\
\midrule
Stage 1 & 100 & 128 & AdamW & 5e-3 & \makecell{5\% W.U + \\ 95\% C.A} & 0.5, 0.5 \\
Stage 2 & 100 & 4 & AdamW & 5e-4 & \makecell{5\% W.U + \\ 95\% C.A} & 0.5, 0.5 \\
\bottomrule
\end{tabular}
\end{table}

\subsection{Training Details}
The dataset is randomly shuffled and partitioned into training, validation, and test sets with an 80:10:10 ratio. The model is optimized via a multi-objective loss function, defined as the weighted sum of Mean Squared Errors (MSE) for the two physical components:
\begin{equation}
    \mathcal{L}_{\text{total}} = \lambda_{\text{PL}}\mathcal{L}_{\text{PL}} + \lambda_{\text{SH}}\mathcal{L}_{\text{SH}},
\end{equation}where $\mathcal{L}_{\text{PL}}$ and $\mathcal{L}_{\text{SH}}$ denote the MSE of the respective components, and $\lambda_{\text{PL}}, \lambda_{\text{SH}}$ are hyperparameters balancing their contributions.
Our training process is divided into two stages to ensure stable convergence.
\begin{itemize}
    \item \textbf{Stage 1: Frozen Image Encoder.} Only the pre-trained image encoder (model card: mobilenetv3\_small\_050) is frozen, and all the other modules of the framework are trained.
    \item \textbf{Stage 2: Fully Fine-tuning.} All modules of the framework are fine-tuned with a lower learning rate.
\end{itemize}
The best configurations for both training stages are summarized in Table~\ref{tab:training-config}. 
\subsection{Evaluation Metrics}
To quantitatively evaluate the performance of our model, we employ three standard metrics for regression tasks. These metrics are applied independently to the predicted RSSI, PL, and SH values. We fairly compare our model with the baseline \cite{arxiv_2025_rgb} since both works were implemented in the same set of DeepSense 6G database \cite{deepsense}. The other outdoor camera-based methods \cite{zhang2025vision,isape_2024_cv} use the same framework as the baseline \cite{arxiv_2025_rgb} yet the datasets are self-collected.
\begin{itemize}
    \item \textbf{Root Mean Squared Error (RMSE)} measures the standard deviation of the prediction errors. It is particularly sensitive to large errors, penalizing them more heavily.
    \item \textbf{Mean Absolute Error (MAE)} measures the average magnitude of the errors, providing a robust and easily interpretable assessment of overall prediction accuracy.
    \item \textbf{1 dB Tolerance Rate (\%Error$\leq$1 dB).} Based on the RSSI absolute error histogram, calculates the percentage of test samples where the absolute prediction error is within 1 dB. This metric indicates the model's consistency and reliability within a practically acceptable tolerance range for wireless applications.
\end{itemize}

\begin{table}[tb]
\centering
\caption{Overall Performance of Our Model}
\label{tab:overall_performance}
\begin{tabular}{llccc}
\toprule
\textbf{Stage} & \textbf{Component} & \textbf{RMSE (dB)} & \textbf{MAE (dB)} & \textbf{\%Error$\leq$1dB} \\
\midrule
\multirow{3}{*}{S1} 
 & PL & 0.67 & 0.41 & 93.25\% \\
 & SH & 0.96 & 0.62 & 81.76\% \\
 & RSSI & 0.96 & 0.63 & 81.50\% \\
\midrule
\multirow{3}{*}{S2} 
 & PL & 0.40 & 0.23 & 97.94\% \\
 & SH & 0.73 & 0.47 & 88.91\% \\
 & RSSI & \textbf{0.70} & \textbf{0.44} & \textbf{89.48\%} \\
\bottomrule
\end{tabular}
\end{table}

\section{Results and Analysis}
In this section, we present a comprehensive evaluation of our proposed framework. We first analyze the overall prediction performance, then conduct two ablation studies to investigate the impact of environmental semantic completeness and the potential of advanced visual encoders. 

\subsection{Overall Performance}
\begin{figure*}[t]
    \centering
    \begin{subfigure}{0.3\textwidth}
        \includegraphics[width=\linewidth]{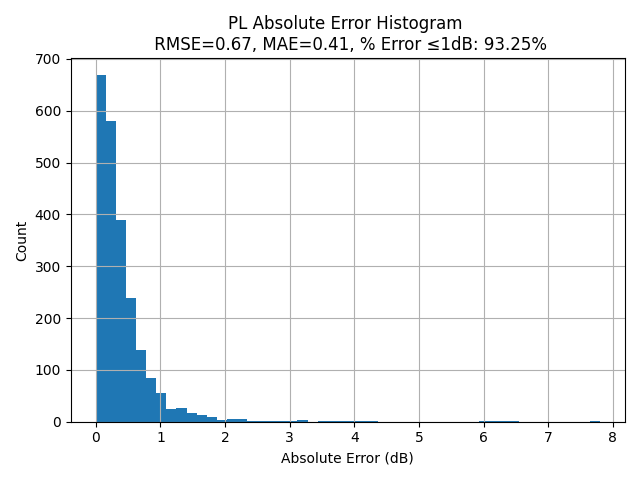}
        \caption{PL (Stage 1)}
        \label{fig:s1_pl_hist}
    \end{subfigure}
    \begin{subfigure}{0.3\textwidth}
        \includegraphics[width=\linewidth]{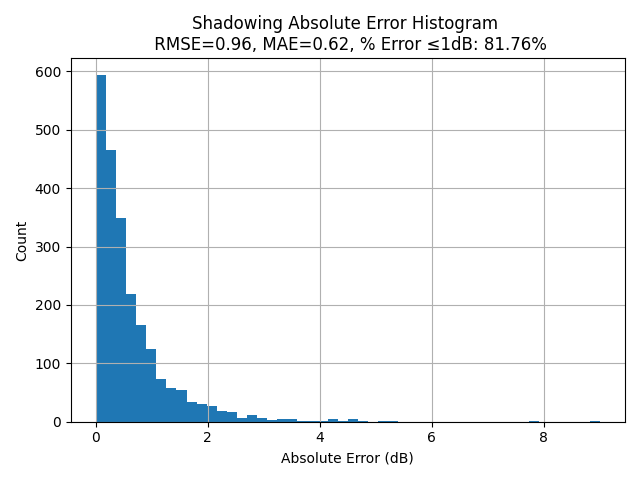}
        \caption{SH (Stage 1)}
        \label{fig:s1_sh_hist}
    \end{subfigure}
    \begin{subfigure}{0.3\textwidth}
        \includegraphics[width=\linewidth]{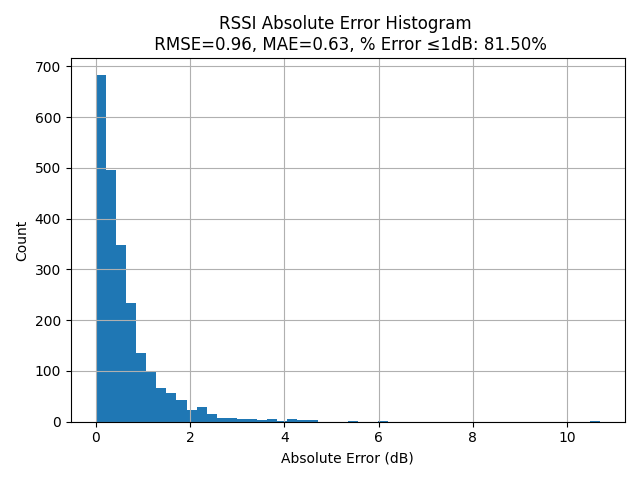}
        \caption{RSSI (Stage 1)}
        \label{fig:s1_rssi_hist}
    \end{subfigure}
    \begin{subfigure}{0.3\textwidth}
        \includegraphics[width=\linewidth]{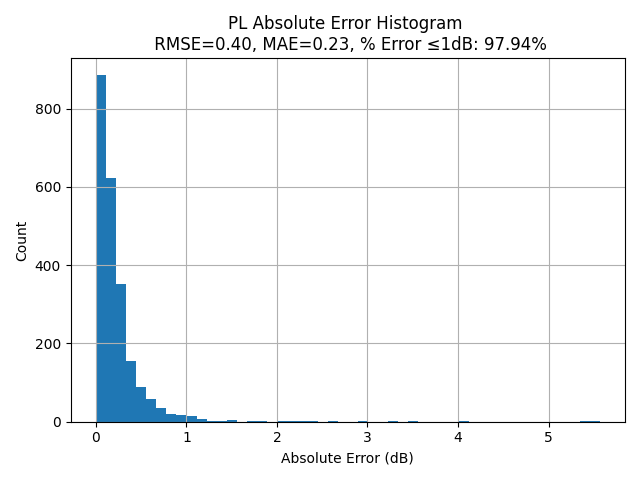}
        \caption{PL (Stage 2)}
        \label{fig:s2_pl_hist}
    \end{subfigure}
    \begin{subfigure}{0.3\textwidth}
        \includegraphics[width=\linewidth]{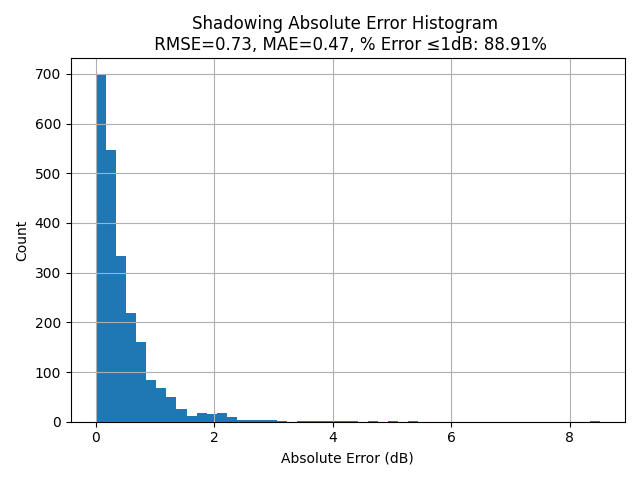}
        \caption{SH (Stage 2)}
        \label{fig:s2_sh_hist}
    \end{subfigure}
    \begin{subfigure}{0.3\textwidth}
        \includegraphics[width=\linewidth]{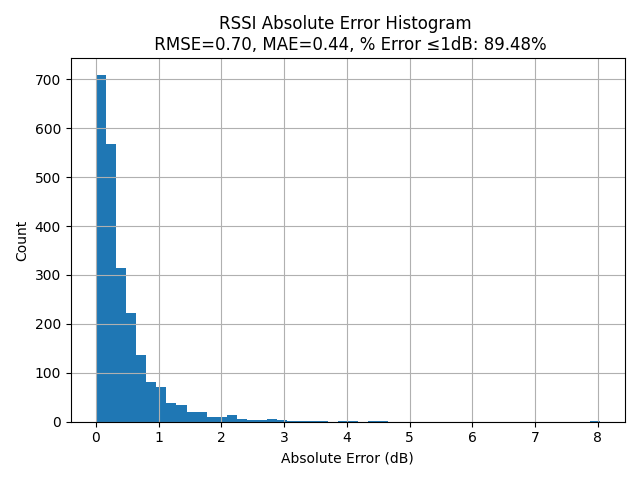}
        \caption{RSSI (Stage 2)}
        \label{fig:s2_rssi_hist}
    \end{subfigure}
    \caption{Absolute error histograms for PL, SH, and RSSI. The top row shows the results after Stage 1 (frozen backbone), and the bottom row shows the improved results after Stage 2 (fully fine-tuning).}
    \label{fig:overall_histograms}
\end{figure*}
Our model demonstrates highly accurate and robust overall performance. This is achieved through a strategic two-stage training process. The results, summarized in \autoref{tab:overall_performance} and visualized in \autoref{fig:overall_histograms}, validate both our framework and our training methodology.

Stage 1 (Frozen Image Encoder) serves as a strong baseline, where only the non-visual components of the model are trained. 
In \autoref{tab:overall_performance}, even with a general-purpose and pre-trained image encoder, the model achieves a respectable RSSI RMSE of 0.96 dB. This initial performance, particularly the accurate PL prediction (RMSE 0.67 dB), demonstrates that our physics-based framework can effectively leverage existing visual features.

The true power of our approach is unlocked in Stage 2 (Full Fine-tuning). By unfreezing the image encoder and fine-tuning the entire model, we observe a significant performance enhancement across all predicted components. The overall RSSI RMSE improves from 0.96 dB to 0.70 dB with 89.48\% of errors falling within 1 dB in this final stage, and SH RMSE drops from 0.96 dB to 0.73 dB. The most dramatic improvement is in PL prediction, where the RMSE is reduced from 0.67 dB to 0.40 dB.

This marked improvement from Stage 1 to Stage 2 proves the value of adapting the visual feature extractor to this specific domain. Fine-tuning allows the encoder to learn features more directly correlated with wireless propagation phenomena, rather than just generic object recognition. As shown in \autoref{fig:overall_histograms}, the error distributions in Stage 2 become even more sharply concentrated around zero, confirming higher precision. Ultimately, these results demonstrate that the synergy between our physics-guided decomposition and a two-stage fine-tuning strategy is key to achieving state-of-the-art accuracy.

\subsection{Comparison to Baseline: Simple and small is the best.}
\begin{table}[tb]
\centering
\caption{Comparison with Baseline \cite{arxiv_2025_rgb} on Performance and Efficiency.}
\label{tab:sota_comparison_detailed}
\begin{tabular}{lcc|c}
\toprule
& \textbf{Ours} & \textbf{Baseline} & \textbf{Gain} \\
\midrule
RMSE (w/o Interf.) (dB) & \textbf{0.70} & 1.41 & \textbf{50.3\% $\downarrow$} \\
RMSE (w/ Interf.) (dB) & \textbf{0.69} & 0.78 & \textbf{11.5\% $\downarrow$} \\
Model Parameters (M) & \textbf{2.51} & 47.7 / 86.1 \textsuperscript{a} & \textbf{19$\times$ / 34$\times$ $\downarrow$} \\
\bottomrule
\end{tabular}
\begin{flushleft}
\small
\textsuperscript{a} The recommended model (47.7 M) and the best-performing model (86.1 M). 
\end{flushleft}
\end{table}

A detailed comparison with the Baseline \cite{arxiv_2025_rgb} reveals that our physically interpretable model is not only more accurate but also vastly more efficient and robust. The results, summarized in Table \ref{tab:sota_comparison_detailed}, support that a simple and small framework, when guided by domain knowledge, is superior to a more complex brute-force approach for this task.

First, our model demonstrates superior robustness to visual interference. Our model's performance remains exceptionally stable, with an RMSE of 0.70 dB when distractors\footnote{For the model training, we keep the bounding box of the Tx and apply zero-padding for the rest of the bounding box inputs.} are present and 0.69 dB when they are removed. In stark contrast, the Baseline model is highly sensitive to such interference. Its RMSE degrades from 0.78 dB to 1.41 dB when interference is not manually eliminated. This highlights the advantage of our approach: by decomposing the prediction into physically meaningful components of path loss and shadow fading, our model learns the underlying propagation principles, making it inherently resilient to noisy visual data. Notably, our model's performance in the presence of interference (0.70 dB) is still better than the Baseline's best performance after eliminating distractors (0.78 dB).

Second, our physics-guided framework leads to a dramatic reduction in model complexity. Our complete end-to-end model consists of only 2.51 M parameters. The Baseline method, a two-stage pipeline, requires 47.7 M (recommended framework) and 86.1 M (highest accuracy framework) parameters, combining a YOLOv8 with a deep ResNet \cite{resnet} (ResNet-34 and ResNet-152). This means our model is approximately 19 to 34 times smaller while delivering higher accuracy and robustness.
This efficiency is a direct result of embedding physical principles into the model's structure, which reduces the reliance on massive parameter spaces to learn the input-output relationship from scratch. Ultimately, our work validates that for channel prediction, a well-structured, lightweight model is not a compromise but an advantage, yielding better performance with a fraction of the computational cost.

\subsection{Impact of Environmental Information}
\begin{table}[tb]
\centering
\caption{Comparison of the prediction with/without the road-sign bounding box.}
\label{tab:error-w-traffic-sign}
\begin{tabular}{llccc}
\toprule
 & \textbf{Metric} & \textbf{PL} & \textbf{SH} & \textbf{RSSI} \\
\midrule
\multirow{3}{*}{w/o} & RMSE & 0.31 & 0.56 & 0.56 \\
 & MAE & 0.23 & 0.42 & 0.42 \\
 & \% Error $\le$1dB & 99.59\% & 93.78\% & 91.70\% \\
\midrule
\multirow{3}{*}{w/} & RMSE & 0.29 & 0.54 & 0.53 \\
 & MAE & 0.22 & 0.41 & 0.40 \\
 & \% Error $\le$1dB & 100.00\% & 94.19\% & 93.36\% \\
\bottomrule
\end{tabular}
\end{table}

We investigate how the inclusion of small but semantically relevant environmental objects, such as traffic signs, affects the accuracy of the prediction. The original bounding box annotations, derived from object detection models, often miss small roadside objects that are visually inconspicuous but physically significant for signal obstruction.

To assess their impact, we manually added the bounding box of the road sign (red bounding box shown in \autoref{fig:framework}) in Scenario 1 of DeepSense 6G  and compared model performance with and without this bounding box. 
Just adding one bounding box which can cause signal obstruction, there is improvement of the prediction. In \autoref{tab:error-w-traffic-sign}, we can notice the improvement of each component prediction and the RSSI prediction. 

The results imply that environmental information strongly correlated with wireless communication is crucial for achieving accurate and interpretable predictions. While large objects (e.g., vehicles, buildings) dominate most spatial occlusion effects, small static objects such as road signs may introduce subtle occlusions that affect fine-grained modeling. Incorporating these elements into environmental features reduces the variance of the error thus enhances the precision of physically interpretable components.

This analysis reinforces the motivation for leveraging richer visual understanding techniques, such as vision-language models (VLMs) \cite{clip}, to capture both large-scale and fine-grained semantics in signal-aware perception.
\section{Conclusion}
In this work, we introduced a physics-guided deep learning framework that sets a new state-of-the-art in wireless signal prediction. By decomposing RSSI into its physical components, our framework reduces RMSE by 50.3\% and demonstrates exceptional robustness to interference, a critical advantage over prior work. This superior performance is delivered by a remarkably lightweight framework: 19 times smaller than the competing framework. The combination of high accuracy, robustness, and low computational cost paves the way for real-time deployment on edge devices.
\bibliographystyle{IEEEtran}
\bibliography{references}

\end{document}